\documentstyle[11pt,twoside,colloqOHP2010,epsfig]{article}
\begin{document}
\title{Planets from the HATNet project}
\author{
	G.~\'A.~Bakos$^1$, 
	J.~D.~Hartman$^1$,
	G.~Torres$^1$,
	G.~Kov\'acs$^2$,
	R.~W.~Noyes$^1$, 
	D.~W.~Latham$^1$, 	
	D.~D.~Sasselov$^1$ \&
	B.~B\'eky$^1$
}
\affil{$^1$ Harvard-Smithsonian CfA, 60 Garden St., 
	Cambridge, MA, USA [gbakos@cfa.harvard.edu]} 
\affil{$^2$ Konkoly Observatory, Budapest}
\begin{abstract} 
We summarize the contribution of the HATNet project to extrasolar
planet science, highlighting published planets (HAT-P-1b through
HAT-P-26b).  We also briefly discuss the operations, data analysis,
candidate selection and confirmation procedures, and we summarize what
HATNet provides to the exoplanet community with each discovery.
\end{abstract}
\section{Introduction} The Hungarian-made Automated Telescope Network
(HATNet; Bakos et al.~2004) survey, has been one of the main
contributors to the discovery of transiting exoplanets (TEPs), being
responsible for approximately a quarter of the $\sim 100$ confirmed
TEPs discovered to date (Fig.~1).  It is a wide-field transit survey,
similar to other projects such as Super-WASP (Pollaco et al.~2006), XO
(McCullough et al.~2005), and TrES (Alonso et al.~2004).  The TEPs
discovered by these surveys orbit relatively {\em bright} stars ($V <
13$) which allows for precise parameter determination (e.g.~mass,
radius and eccentricity) and enables follow-up studies to characterize
the planets in detail (e.g.~studies of planetary atmospheres, or
measurements of the sky-projected angle between the orbital axis of the
planet and the spin axis of its host star).  Since 2006, HATNet has
announced twenty-six TEPs\footnote{Meaning that the scientific results were
submitted to peer reviewed journals and posted to
arXiv, the planet host stars have been uniquely identified,
and all discovery data have been made public.}.  Below we highlight some
of the exceptional properties of these planets (Section~2), we then
describe the procedures which we followed to discover them (Section~3),
and we conclude by summarizing what HATNet provides to the TEP
community with each discovery (Section~4).

\begin{figure}[ht]
\begin{center}
\epsfig{width=10cm,file=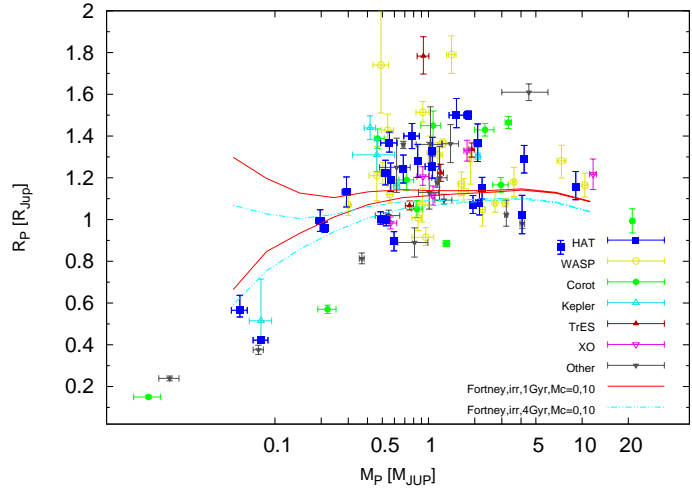}
\caption{
Mass--radius relation of TEPs, highlighting the findings from different
surveys.  The dotted lines are lines of constant density from
0.4\,g\,cm$^{-3}$ to 11.9\,g\,cm$^{-3}$. Also overlaid are models from
Fortney et al.~2007.
}
\end{center}
\end{figure}

\section{Highlights}

HATNet-detected TEPs span a wide range of physical properties,
including: two Neptune-mass planets (HAT-P-11b, Bakos~et~al.~2010a; and
-26b, Hartman~et~al.~2010b); planets with masses greater than $7~M_{\rm
J}$ (-2b, Bakos et al.~2007; and -20b, Bakos~et~al.~2010b); compact
planets (-2b, and -20b); inflated planets (-7b, P\'al~et~al.~2008; -8b,
Latham et al.~2009; -12b, Hartman et al.~2009; -18b, and -19b, Hartman
et al.~2010a); a planet with a period of just over one day (-23b, Bakos
et al.~2010b); planets with periods greater than 10 days (-15b,
Kov\'acs et al.~2010; and -17b, Howard et al.~2010); multi-planet
systems (-13b,c, Bakos et al.~2009; and -17b,c); and a number of
eccentric planets (-2b; -11b; -14b, Torres et al.~2010; -15b; -17b; and
-21b, Bakos et al.~2010b).  We have also provided evidence for outer
planets for 4 systems: HAT-P-11c, -13c, -17c (the latter two with
almost closed orbits), and HAT-P-19c.

Some of these discoveries were the first of their kind, and thus were
important landmarks in exoplanet science.  This includes: the first
transiting heavy-mass planet (-2b); the first retrograde planet (-7b;
Narita et al.~2009, Winn et al.~2009); two of the first four transiting
Neptunes; the first inflated Saturn (-12b); the first and second
multi-planet systems with transiting inner planets; and two of the
first six planets with periods longer than 10 days.

\section{How HATNet Discovers Planets}

The 26 HATNet TEPs were identified from a shortlist of 1300
hand-selected transit {\em candidates} culled from millions of light
curves, which were, in turn, the result of diverse activities ranging
from remote hardware operations to data analysis.  Here we briefly
describe this process.

\subsection{Instrumentation}

HATNet utilizes 6 identical instruments, each with an 11\,cm aperture
f/1.8 lens and a $\rm 4K\times 4K$ front-illuminated CCD with 9\arcsec\
pixels (yielding a wide, $10.6^{\circ}\times10.6^{\circ}$ field),
attached to a horseshoe mount, protected by a clam-shell dome, and with
all devices controlled by a single PC\@.  Each instrument, called a HAT
(Bakos~et~al.~2002), can obtain per-image photometric precision
reaching 4\,mmag at 3.5-min cadence on the bright end at $r\approx
9.5$, and 10\,mmag at $r\approx 12.1$.  By collecting a light curve
with $\sim 100$ or more points in transit, a transit with a depth of
only a few mmag may be detected.  We note that the original HATNet
hardware employed $\rm 2K\times 2K$ front illuminated detectors with
Cousins $I$-band filters.  This was replaced to $\rm 4K\times 4K$
front-illuminated CCDs and Cousins $R$ filters in 2007 September, and
the filter was changed to Sloan $r$ in 2008 July.

Four HAT instruments are located at the Smithsonian Astrophysical
Observatory's (SAO) Fred Lawrence Whipple Observatory (FLWO), and an
additional two instruments are on the roof of the hangar servicing the
antennae of SAO's Submillimeter Array, at Mauna Kea Observatory (MKO)
in Hawaii.  The network with its current longitude coverage has
significant advantages in detecting TEPs with periods longer than a few
days.

\subsection{Data acquisition and TEP Candidate Selection}

The instruments are truly autonomous in the sense that they are aware
of the observing schedule and the weather conditions, they prepare all
the devices (CCDs, dome, telescope) for the observations, acquire ample
calibration frames (biases, darks, skyflats), and then proceed to the
science program of the night.  For the purpose of monitoring bright
stars for transits, the sky has been split up to 838 $8^{\circ} \times
8^{\circ}$ non-overlapping fields.  Fields are chosen for observation
based on several factors such as optimal visibility at the given time
of the year, proximity of the field to Solar system objects, and
various other factors.  To date HATNet has observed $\sim 120$ fields
(29\% of the northern sky).  Typically a field is monitored for 3
months; a given instrument will begin observations of the field after
evening twilight and observe it continuously at a cadence of 3.5
minutes until the field sets.  The instrument will then target a second
field and continue observing it until morning twilight.  All time
between dusk and dawn is spent by exposing on the selected fields.  A
single field is typically assigned to a FLWO instrument as well as a
MKO instrument to increase the duty cycle of the observations.  Based
on operations since 2003, we find that the effective duty cycle of
HATNet is $\sim 38\%$.

The images are calibrated using standard techniques that take into
account the problems raised by the wide FOV, such as strong vignetting,
distortions, sky-background changes, etc.  The entire data flows to the
CfA via fast Internet.  The astrometric solution is determined
following P\'al \& Bakos~(2006), based on the Two Micron All Sky Survey
(2MASS; Skrutskie~et~al.~2006) catalog.  We then perform aperture
photometry at the fixed positions of the 2MASS stars.  An initial
ensemble magnitude calibration is performed, and the remaining
systematic variations are removed from the light curves by
decorrelating against external parameters (i.e.~parameters describing
the point spread function, the position of the star on the image, and
others) and by making use of the Trend Filtering Algorithm (TFA;
Kov\'acs~et~al.~2005).  The resulting light curves typically reach a
per-point photometric precision (at 3.5\,min cadence) of $\sim 4$\,mmag
for the brightest non-saturated stars.

We search the trend filtered light curves for periodic transit events
using the Box-fitting Least Squares algorithm (BLS;
Kov\'acs~et~al.~2002).  We then subject potential transit candidates to
a number of automatic filters to select reliable detections which are
consistent with a transiting planet-size object, and are not obviously
eclipsing binary star systems or other types of variables.

\subsection{Selecting, Confirming, and Characterizing TEPs}

To go from the lengthy list of potential transit candidates to a much
smaller number of confirmed and well-characterized TEPs, we first
manually select a list of candidates to consider for follow-up
observations, we gather and analyze spectroscopic and photometric
observations to reject false positives and confirm bona fide TEPs, and
we finally analyze and publish the confirmed TEP systems.

The automated transit candidate selection procedure provides a
manageable list of potential candidates which must then be inspected by
eye to select the most promising targets for follow-up.  Typically a
few hundred to one thousand potential candidates per field are
identified by the automated procedures, which are then narrowed down to
a few dozen candidates deemed worthy of follow-up.  At the end of this
procedure, relative priorities are assigned to these candidates and
appropriate facilities for follow-up are identified.

Candidates selected for follow-up then undergo a procedure consisting
of three steps: reconnaissance spectroscopy, photometric follow-up
observations, and high-precision spectroscopy.

We have found that a very efficient method to reject the majority of
false positive transit detections is to first obtain one or more
high-resolution, low signal-to-noise ratio (S/N$ \sim 25$ per
resolution element) spectra using a $1$--$2$\,m telescope
(e.g.~Latham~et~al.~2009).  Double-lined eclipsing binaries, giant
stars (where the detected transit is most likely a blend between an
eclipsing binary and the much brighter giant), and rapidly rotating
stars or excessively hot stars (where confirming a planetary orbit
would be very difficult) may be immediately rejected based on one
spectrum.  Stars that are confirmed to be slowly rotating dwarfs are
observed a second time at the opposite quadrature phase to look for a
significant velocity variation.  Stars where the RV amplitude implies a
stellar companion (typically $\ga 5$\,km\,s$^{-1}$) are rejected.  In
some cases changes in the shape of the spectral line profile may also
be detected, enabling us to reject the target as a stellar triple.

We find that a significant fraction of the initial candidates selected
for follow-up are rejected by this reconnaissance spectroscopy (RS)
procedure, saving time on precious resources that we use for the final
follow-up.  To carry out the RS work we primarily use the FLWO~1.5\,m
telescope, previously with the Digital Speedometer (DS), and now with
the Tillinghast Reflector Echelle Spectrograph (TRES), and to some
extent the FIber-fed Echelle Spectrograph (FIES) on the Nordic Optical
Telescope (NOT) at La Palma.  We have also made use of the echelle
spectrograph on the Du Pont 2.5\,m telescope at LCO, the echelle
spectrograph on the ANU 2.3\,m telescope at SSO, and the CORALIE
Echelle spectrograph on the 1.2\,m Euler telescope at La Silla
Observatory (LSO) in Chile.

If a candidate passes the RS step, we then schedule photometric
observations of the candidate over the course of a transit to confirm
that the transit is real, and to confirm that the shape of the transit
light curve is consistent with a TEP.  We also note that these
follow-up light curves are essential for obtaining precise measurements
of system parameters, such as the planet-star radius ratio or the
transit impact parameter.  In some cases the candidate is subjected to
photometry follow-up without RS or with incomplete RS, when, e.g., there
is evidence that the star is a dwarf (colors, proper motion, parallax),
or when the transit events are rare (long period).  We primarily use
the KeplerCam instrument on the FLWO~1.2\,m telescope, but have also
made use of Faulkes Telescope North (FTN) of the Las Cumbres
Observatory Global Telescope (LCOGT) on Mauna Haleakala, Hawaii, and
occasionally telescopes in Hungary and Israel.

The final step in the confirmation follow-up procedure is to obtain
high-resolution, high-S/N spectra with sufficient velocity precision to
detect the orbital variation of the star due to the planet, confirm
that the system is not a subtle blend configuration, and measure the
effective temperature, surface gravity and metallicity of the star.  By
this stage we have already excluded the majority of false positives, so
that roughly half of the candidates that reach this step are confirmed
as TEPs.  False positives that reach this stage generally are rejected
after only a few spectra are obtained, so that $\sim 75\%$ of the time
is spent observing TEPs.  We primarily use the HIgh Resolution Echelle
Spectrometer (HIRES) with the iodine cell on the 10\,m Keck I telescope
at MKO.  We have also used the FIES/NOT facility, the High Dispersion
Spectrograph (HDS) with the iodine cell on the Subaru 8.2\,m telescope
at MKO, and the SOPHIE instrument on the 1.93\,m telescope at the
Observatoire de Haute-Provence (OHP), in France.

Once a planet is confirmed, we conduct a joint analysis of the
available high-precision RV observations and photometric observations
to determine the system parameters, including in particular the masses
and radii of the star and planet(s) (e.g.~Bakos~et~al.~2010a).  In
cases where the spectral line bisector spans are inconclusive, we must
also carry out a detailed blend-model of the system concurrently with
the TEP modeling to definitively prove that the object is a TEP.

%
\section{What HATNet Provides}

The HATNet project strives to provide the following to the community
for a given TEP discovery:

\begin{enumerate}
\item HATNet discovery data.
\item High-precision, often multi-band, photometry follow-up.
\item High-precision radial velocity follow-up.
\item Access to all of the data via the online tables, including raw
	and detrended values.
\item Characterization of the host star: stellar atmospheric and
	fundamental parameters (from isochrone fitting).
\item A blend analysis.
\item Accurate parameters of the planetary system.
\item A publication available on arXiv with all the above 
when the planet is announced.
\end{enumerate}

\acknowledgments{
	HATNet operations have been funded by NASA grants
	NNG04GN74G, NNX08AF23G and SAO IR\&D grants. Work of G.\'A.B. was
	supported by the Postdoctoral Fellowship of the NSF Astronomy and
	Astrophysics Program (AST-0702843). We wish to thank all our
	collaborators the help they have provided over the past years.
}

\end{document}